\documentclass{rnaastex}

\begin{document}

\title{Ghosts of Jupiter's past: is 2017~UV$_{43}$ a relative of comet Shoemaker-Levy 9?}

\correspondingauthor{Carlos~de~la~Fuente~Marcos}
\email{nbplanet@ucm.es}

\author[0000-0003-3894-8609]{Carlos~de~la~Fuente~Marcos}
\affiliation{Universidad Complutense de Madrid \\
             Ciudad Universitaria, E-28040 Madrid, Spain}

\author[0000-0002-5319-5716]{Ra\'ul~de~la~Fuente~Marcos}
\affiliation{Universidad Complutense de Madrid \\
             Ciudad Universitaria, E-28040 Madrid, Spain}


\section{} 

The first collision of two Solar System bodies ever to be observed was that of the fragments of comet D/1993~F2 (Shoemaker-Levy~9) with Jupiter 
from July 16 through July 22, 1994 (see e.g. \citealt{1994A&A...289..607S,1998P&SS...46...21S}). Backward integrations showed that Shoemaker-Levy~9 
\citep{1993IAUC.5725....1S} may have become an eccentric Jovian satellite in 1929$\pm$9, having been torn into at least 21 major pieces during a 
very close encounter with Jupiter in July 1992 (see e.g. \citealt{1994A&A...289..607S,1998P&SS...46...21S,1995Icar..118..155B}).

Here, we present numerical evidence suggesting that 2017~UV$_{43}$ might be related to Shoemaker-Levy~9. With our exploratory results we intend to 
encourage a search for precovery images of 2017~UV$_{43}$ and perhaps even follow-up observations that may help in improving its poorly determined 
orbit so its true dynamical nature is understood. The current orbit determination of 2017~UV$_{43}$ (epoch JD~2458000.5, 4-September-2017, instant 
zero of time) is based on 15 observations for a data-arc span of 10~d and has semi-major axis, $a$=6.63$\pm$0.14~au, eccentricity, $e$=0.21$\pm$0.07, 
inclination, $i$=5\fdg20$\pm$0\fdg04, longitude of the ascending node, $\Omega$=321\fdg0$\pm$1\fdg0, and argument of perihelion, 
$\omega$=181\degr$\pm$13\degr; with an absolute magnitude of 14.0$\pm$0.4 it may be 3--9~km 
wide.\footnote{\href{http://ssd.jpl.nasa.gov/sbdb.cgi}{JPL's Small-Body Database}} The average values and standard deviations of the orbital elements 
of the 21 fragments of Shoemaker-Levy~9 with orbit determinations are: $a$=6.81$\pm$0.04~au, $e$=0.210$\pm$0.004, $i$=5\fdg87$\pm$0\fdg08, 
$\Omega$=220\fdg8$\pm$0\fdg2, and $\omega$=354\fdg98$\pm$0\fdg05. Their estimated sizes range from $\sim$0.1 to $\sim$4~km 
\citep{1995A&A...304..296S,2004SoSyR..38..219Z}. 

\begin{figure}[!ht]
\begin{center}
\includegraphics[scale=0.35,angle=0]{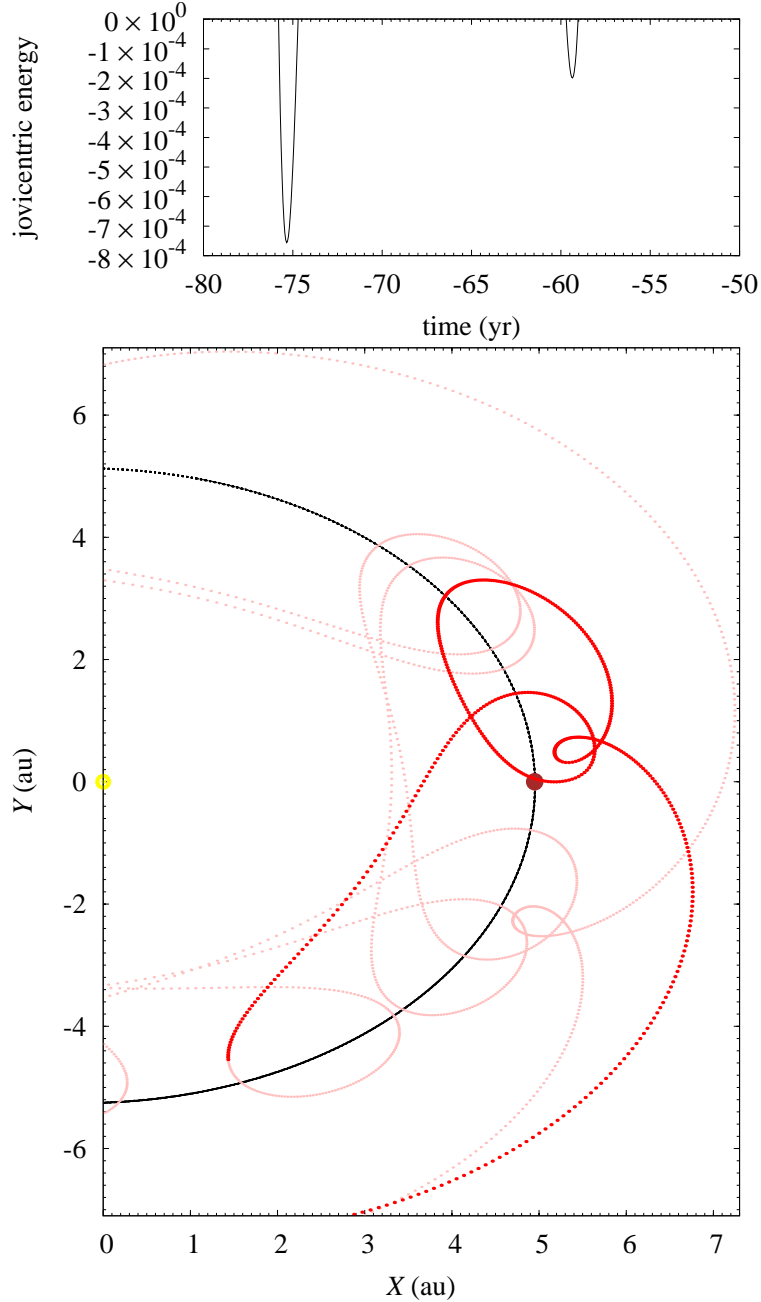}
\caption{Evolution of the Keplerian jovicentric energy of 2017~UV$_{43}$ (top panel). Satellite captures happen when the relative binding energy 
         becomes negative. The unit of energy is such that the unit of mass is 1~$M_{\odot}$, the unit of distance is 1~au and the unit of time 
         is one sidereal year divided by 2$\pi$. The path followed by 2017~UV$_{43}$ (that moves clockwise) in a frame of reference centered at 
         the Sun (in yellow) and rotating with Jupiter (in brown, its orbit in black), projected on to the ecliptic plane, during the time 
         interval ($-$200, 0)~yr is showed in pink (bottom panel), with the interval displayed in the top panel plotted in red.  
\label{fig:1}}
\end{center}
\end{figure}

The present-day orbit of 2017~UV$_{43}$ resembles ---$a$, $e$ and $i$--- the average one of the fragments of Shoemaker-Levy~9, but it is unclear 
whether such similarity can be extended to the backward dynamical evolution. We have used the orbit determination of 2017~UV$_{43}$ to perform a 
preliminary study of its short-term evolution (for technical details, see \citealt{2012MNRAS.427..728D}). The results of our limited analysis 
based on the rather uncertain orbital solution strongly suggest that its dynamical evolution becomes difficult to reconstruct beyond a few decades 
into the past. Figure~\ref{fig:1}, top panel, shows that the binding energy relative to Jupiter of 2017~UV$_{43}$ became negative during two 
recent flybys (nearly 75.3 and 59.3~yr ago, i.e. in 1942.38 and 1958.38) for about one sidereal year. The flybys took place after the capture of 
Shoemaker-Levy~9 by Jupiter. The path followed by 2017~UV$_{43}$ during the relevant time interval is displayed in Figure~\ref{fig:1}, bottom 
panel. Although both the timing and the dynamics may be consistent by chance, 2017~UV$_{43}$ might be a lost fragment of Shoemaker-Levy~9. We have 
gathered the available astrometry of 2017~UV$_{43}$ and Shoemaker-Levy~9, but we have been unable to find a satisfactory orbit determination 
capable of fitting both data sets. Therefore, we must assume that, if 2017~UV$_{43}$ is genetically related to the fragments of Shoemaker-Levy~9, 
their orbital connection was severed long before 1992. 

Asteroid 2017~UV$_{43}$ may be as large as the largest observed fragments of Shoemaker-Levy~9. If there is a physical connection between them, 
comet Shoemaker-Levy~9 as first observed in 1993 might not have been primordial in the sense that the breakup experienced in 1992 might not 
have been its first. Conversely, assuming no physical connection between Shoemaker-Levy~9 and 2017~UV$_{43}$, the orbital similarity may signal 
the existence of an active delivery route placing minor bodies into Shoemaker-Levy~9-like orbits. Follow-up observations of 2017~UV$_{43}$ during 
the next year will not be difficult. It will reach its next opposition late in November 2018 at $V$=21.3~mag. This will be the best opportunity to 
acquire spectroscopy that may help in understanding its origin.

\acknowledgments

We thank S.~J. Aarseth for providing the code used in this research and A.~I. G\'omez de Castro of the Universidad Complutense de Madrid (UCM) for 
providing access to computing facilities. This work was partially supported by the Spanish MINECO under grant ESP2014-54243-R. Part of the 
calculations and the data analysis were completed on the EOLO cluster of the UCM. EOLO is funded by the MECD and MICINN. This is a contribution 
to the CEI Moncloa. In preparation of this Note, we made use of the NASA Astrophysics Data System and the MPC data server.

\end{document}